\include{graphicsx}

\documentclass[12pt,preprint]{aastex}

\slugcomment{Submitted Oct 30, 2001; accepted Jan 23, 2002, to be published in March 2002 ApJ Letters}

\shorttitle{Survey for M8-M9 binaries}
\shortauthors{Close et al.}

\begin{document}


\title{An Adaptive Optics Survey of M8-M9 stars: Discovery of 4 Very Low Mass Binaries With at Least One System Containing a Brown Dwarf Companion}


\author{Laird M. Close$^1$, Nick Siegler$^1$, Dan Potter$^2$, Wolfgang Brandner$^3$, James Liebert$^1$}

\email{lclose@as.arizona.edu}

\affil{$^1$Steward Observatory, University of Arizona, Tucson, AZ 85721}
\affil{$^2$Institute for Astronomy, University of Hawaii, Honolulu, HI}
\affil{$^3$European Southern Observatory, Garching, Germany}   



\begin{abstract} 

Use of the highly sensitive Hokupa'a/Gemini curvature wavefront sensor has
allowed for the first time direct adaptive optics (AO) guiding on
M8-M9 very low mass (VLM) stars. An initial survey of 20 such objects
(SpT=M8-M9) discovered 4 binaries. Three of the systems have
separations of less than 4.2 AU and similar mass ratios ($\Delta
K<0.8$ mag; $0.85<q<1.0$).  One system, however, did have the largest
$\Delta K=2.38$ mag and $sep=14.4$ AU yet observed for a VLM star with
a brown dwarf companion. Based on our initial flux limited ($Ks<12$
mag) survey of 20 M8-M9 stars over $14:26<RA<4:30$ hours from the
sample of \cite{giz00} we find a binary fraction in the range
$14-24\%$ for M8-M9 binaries with $sep>3$ AU. This is likely
consistent with the $23\pm 5\%$ measured for more massive (M0-M6)
stars over the same separation range. It appears M8-M9 binaries have a
much smaller semi-major axis distribution peak ($\sim 4$ AU; with no
systems wider than 15 AU) compared to M and G stars which have a broad
peak at larger $\sim 30$ AU separations.



\end{abstract}


\section{Introduction}

Since the discovery of Gl 229B by \cite{nak95} there has been intense
interest in the direct detection of brown dwarfs and very low mass
(VLM) stars. According to the current models of \cite{bur00} and \cite{cha01}, stars
with spectral types of M8-M9 will be just above the stellar/substellar
boundary. However, most fainter companions to such primaries should
themselves be substellar. Therefore, a survey of M8-M9 stars should
detect binary systems with VLM primaries with VLM or brown dwarf
secondaries.

The binary frequency of M8-M9 stars is interesting in its own right,
since little is known about how common M8-M9 binary systems are. It is
not clear currently if the M8-M9 binary separation distribution is
similar to that of M0-M6 stars; in fact, there is emerging evidence
that very low mass L dwarf binaries tend to have smaller separations
but similar binary frequencies as more massive M and G stars (\cite{mar99} \&
\citep{rei01a}). 



In this letter we present 3 newly discovered M8-M9 binaries (2MASSW
J2140293+162518, 2MASSWJ 2206228-204705, and 2MASSW J2331016-040618
--hereafter 2M2140, 2M2206 and 2M2331, respectively). Earlier in this
survey we discovered another M9 binary 2MASSJ 1426316+155701
(hereafter 2M1426) for which a detailed description was already
published in \cite{clo02}. However, we have re-analyzed the data from
\cite{clo02} and include 2M1426 here for completeness since we have revised the mass estimate for this system. 

These 4 new binaries are a significant addition to the $\sim 9$ very
low mass binaries known to date (\cite{rei01a} \& \cite{clo02}). With
relatively short periods these new systems will likely play a
significant role in the mass luminosity calibration for VLM stars and
brown dwarfs. It is also noteworthy that we can start to characterize
this new population of M8-M9 binaries. We will outline how M8-M9
binaries are both similar and different from their more massive
M \& G counterparts.



\section{An AO survey of nearby M8-M9 field stars}

 As outlined in detail in
\cite{clo02} we utilized the University of Hawaii curvature adaptive optics (AO) system
Hokupa'a (Graves et al 1998; Close et al. 1998) which is a visitor AO
instrument on the Gemini North Telescope. This highly sensitive
curvature AO system is well suited to locking onto nearby, faint, red
M8-M9 stars and producing $0.13\arcsec$ images (which are close to the
$0.07\arcsec$ diffraction-limit in the $K^\prime$ band). We utilized this unique capability to survey the nearest extreme M stars
(M8.0-M9.5) to characterize the nearby M8-M9 binary
population.

Here we report the results of our second observing run on UT Sept 22,
2001. We targeted M8-M9 VLM stars identified by
\cite{giz00}. We have observed 20 out of 24 of the published
 \citep{giz00} M8.0-M9.5 stars with $Ks<12$ mag over the $RA$ range
 14:26-04:30 hours. The four systems not observed due to time
 constraints were 2MASSW J145739+451716, J1553199+140033,
 J1627279+810507, J2349489+122438, all other M8-M9 stars with $Ks<12$
 in the $RA$ range 14:26-04:30 hours have been observed from the list
 of \cite{giz00}. It should be noted that the M8-M9 list of
 \cite{giz00} has some selection constraints: galactic latitudes are
 all $> 20$ degrees; from $0<RA<4.5$ hours $DEC<30$ degrees; and there are
 gaps in the coverage due to the past availability of the 2MASS scans.


Four of our 20 targets were clearly tight binaries
($sep < 0.5 \arcsec$). We observed each of these objects by dithering
over 4 different positions on the QUIRC NIR 1024x1024 detector with
$0.0199\arcsec$/pix (\cite{hod96}). At each position we took 3x10s exposures at
J, H, $K^{\prime}$, and 3x60s exposures at H. Resulting in unsaturated
120s exposures at J, H, and $K^{\prime}$ with deep 720s exposure
at H band for each binary system.

\section{Reductions \& Analysis}

We have developed an AO data reduction pipeline in the IRAF language
which maximizes sensitivity and image resolution. This pipeline is standard IR AO data reduction and is described in detail in \cite{clo02}. 

\begin{figure}
\includegraphics[angle=0,width=\columnwidth]{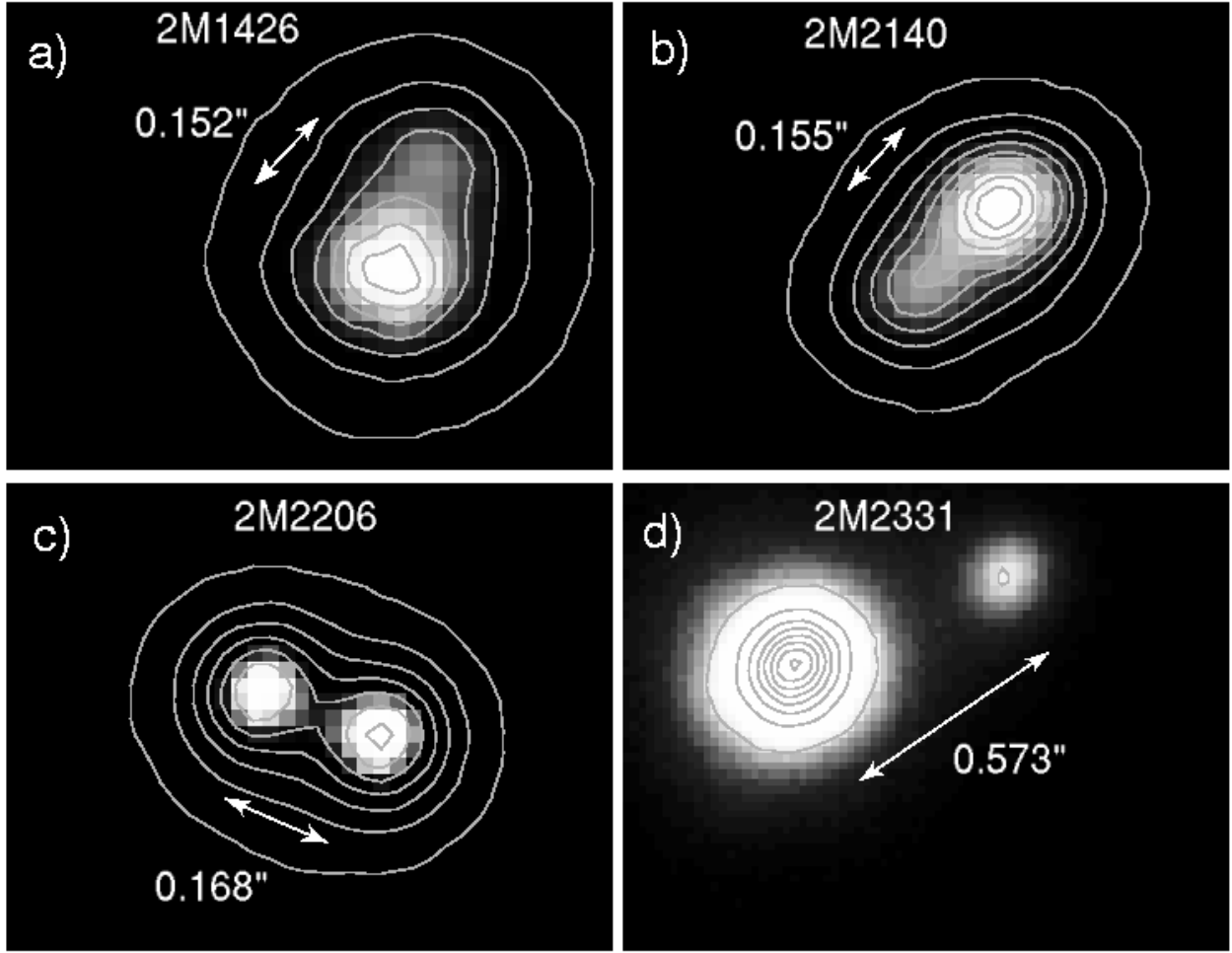}\caption{In figure (a) we see the 12x10s $K^\prime$ image of the 2MASSJ 1426316+155701 binary discussed in \cite{clo02} at a resolution of
$0.131\arcsec$. In Figure (b-d) we show $K^\prime$ images of the new binaries 2MASSW J2140293+162518, 2MASSWJ 2206228-204705, and 2MASSWJ 2331016-040618, respectively. The pixels are
0.0199$\arcsec$/pix. The contours are linear at the 90, 75, 60, 45, 30, 15, and
1\% levels. North is up and East left in each figure.\label{fig1}} \end{figure}

This pipeline produced final unsaturated 120s exposures at J ($FWHM
\sim 0.15\arcsec$) , H ($FWHM \sim 0.14 \arcsec$), and $K^{\prime}$
($FWHM \sim 0.13 \arcsec$) with a deep 720s exposure ($FWHM \sim 0.14
\arcsec$) at H band for each binary system. The dithering produces a
final image of 30x30$\arcsec$ with the most sensitive region
(10x10$\arcsec$) centered on the binary. See Figure
\ref{fig1} which illustrates $K^{\prime}$ images of each of the new systems.
We made the small conversion from the $K^\prime$ magnitudes to K with
the calibration of $K-K^\prime=0.22(H-K)$ 
which was derived for similarly reddened stars \citep{wai92}.

\placefigure{fig1} 

In Table \ref{tbl-1} we present the analysis of the images taken of
the 4 new binaries from both runs. The photometry was based on DAOPHOT
 PSF fitting photometry \citep{ste87}. The PSF used was the reduced
12x10s unsaturated data from the next (single) brown dwarf observed
after each binary. The PSF "star" always had a similar IR brightness, a late M spectral type, and was observed at a similar
airmass.  The resulting $\Delta magnitudes$ are listed in Table
\ref{tbl-1}. The errors in $\Delta mag$ are the differences in the photmetry between 2 similar PSF stars. The individual fluxes were calculated from the flux ratio measured by DAOPHOT (assuming $\Delta K' = \Delta Ks$) and the
total flux of the binary in a $15\arcsec$ aperture scaled to the
published 2MASS fluxes of the blended binary.

The platescale and orientation of QUIRC was determined from a short
exposure of the Trapezium cluster in Orion and compared to published
positions as in \cite{sim99}. From these observations a platescale of
$ 0.0199\pm 0.0002\arcsec$/pix and an orientation of the Y-axis
($0.3\pm 0.3$ degrees E of north) was determined. The astrometry for
each binary was based on the PSF fitting. The astrometric errors are
based on the range of values observed at the different wavelengths and
the systematic errors in the calibration added in quadrature.


\placetable{tbl-1}

\placetable{tbl-2}

\section{Discussion}

\subsection {Are the companions physically related to primaries?}


Since \cite{giz00} only picked objects $>20$ degrees above the
galactic plane we do not expect many background late M or L
stars in our images. In the $1.8x10^4$ square arcsecs already surveyed, we
have not detected a $J-Ks>0.8$ mag background object in any of the
fields. Therefore, we estimate the probability of a chance projection
of such a red object within $<0.5\arcsec$ of the primary is
$<4$x$10^{-5}$. We conclude that all these very red, cool objects are
physically related to their primaries and hereafter refer to them as
2M1426B, 2M2140B, 2M2206B and 2M2331B.

\subsection {What are the spectral types of the components?}

We do not have spatially resolved spectra of both components in any of
these systems; consequently we can only try to fit the observed $J-Ks$
colors in Table \ref{tbl-2} to spectral templates. Unfortunately, the
exact relationship between IR colors and brown dwarf spectral types is
still under study. However, according to the observations of
\cite{rei01b} our observed $J-Ks$ colors can best be fit by the spectral
types in the $7^{th}$ column of Table \ref{tbl-2}. It is important to
note that these spectral types are only a guide since the conversion
from $J-Ks$ to spectral type carries at least $\pm1.5$ spectral subclasses of
uncertainty. Fortunately none of the following analysis is dependent on these spectral type estimates.


\subsection {What are the distances to the binaries?}

\begin{figure}
\includegraphics[angle=270,width=250pt]{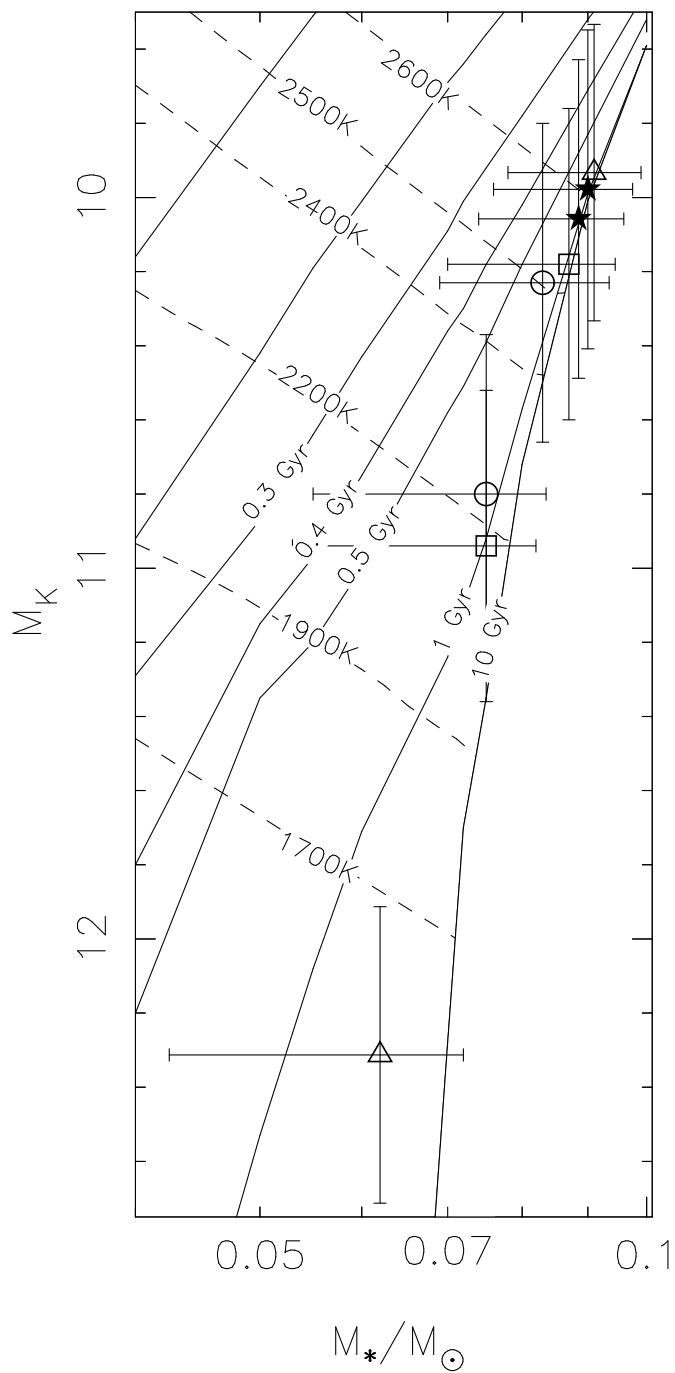}\caption{The latest \cite{cha01} DUSTY evolutionary models. The locations of the 2 components of 2M1426 are
indicated by the circles, 2M2140 squares, 2M2206 solid stars, and 2M2331 by the triangles. The large error bars are due to the errors in the distance, photometry, and age of the systems $0.6-7.5$ Gyr. The $M_K$ of each secondary is determined by the addition of $\Delta K$ plus the $M_K$ of the primary. Note that 2M2331 has the largest mass ratio (q=0.68) of any M8 binary known to date. \label{fig2}} \end{figure}

Unfortunately, there are no published trigonometric parallaxes to any
of these systems. We can estimate, however, the distance based on the
trigonometric parallaxes of other M8-M9 stars. The distances of all the primaries were
determined from the absolute K magnitudes which can be estimated by
$M_K=7.593+2.25(J-Ks)$ for M8-M9 stars
\citep{giz00}. This relationship has a $1 \sigma$ error of 0.36 mag
which has been added in quadrature with the J \& Ks photometric errors
to yield the primary component's $M_K$ values plotted in Figure \ref{fig2}.



\placefigure{fig2}

\subsection {What are ages of the systems?}

Estimating the exact age for any of these systems is difficult since
there are no Li measurements yet published (which could place an upper
limit on the ages). For completeness we have assumed the whole range
of common ages in the solar neighborhood (0.6-7.5 Gyr) may apply to each
system \citep{cal99}. However, \cite{giz00} observed very low proper motion ($V_{tan}<10$ km/s) for the
2M2140 and 2M2206 systems.
These two systems are among the lowest velocity M8's in the entire survey of
\cite{giz00}. This suggests a somewhat younger age since these systems have
not yet developed a significant random velocity like the other older ($\sim5$ Gyr) M8-M9
stars in the survey. 
Therefore, we assign a slightly younger age of $3.0^{+4.5}_{-2.4}$ Gyr
to these 2 systems (2M2140 and 2M2206), but leave large error bars
allowing ages from 0.6-7.5 Gyr ($\sim 3$ Gyr is maximum age for the kinematically young stars found by \cite{cal99}).  The other binary system 2M2331
appears to have a normal $V_{tan}$ and so is more likely to be an
older system. Hence we assign an age of $5.0_{-4.4}^{+2.5}$ Gyr which is an average age for a star in the solar neighborhood \citep{cal99}.

\subsection {The Masses of the Components}

To estimate masses for these objects we will need to rely on theoretical evolutionary tracks for VLM stars and brown dwarfs. Calibrated
theoretical evolutionary tracks are required for objects in the
temperature range 1600-2600 K. Recently such a calibration has been
performed by two groups using dynamical measurements of the M8.5 Gl569B
brown dwarf binary. From the dynamical mass measurements of the Gl569B
binary brown dwarf (\cite{ken01}, and \cite{lan01}) it was found that
the \cite{cha01} and \cite{bur00} evolutionary models were in reasonably
good agreement with observation.  In Figure
\ref{fig2} we plot the latest DUSTY
models from \cite{cha01}.

\subsection {The masses of the components}


We can estimate the masses of the components based on the age range of
0.6-7.5 Gyr and the range of $M_K$ values. The maximum mass relates to
the minimum $M_K$ and the maximum age of 7.5 Gyr. The minimum mass
relates to the maximum $M_K$ and the minimum age of 0.6 Gyr. These
masses are listed in Table
\ref{tbl-2} and illustrated in Figure \ref{fig2}. 

At the younger ages ($<1 Gyr$), the primaries may be on the
stellar/substellar boundary, but they are most likely VLM stars. The
substellar nature of the companion is very likely in the case of
2M2331B, possible in the cases of 2M1426B and 2M2140B, to unlikely in
the case of 2M2206B which appears to be a VLM star like its
primary. Hence 3 of the companions may be
brown dwarfs.

\subsection {The binary frequency of M8-M9 stars}

We have carried out a flux limited ($Ks<12$) survey of 20 M8-M9
primaries. Around these 20 M8-M9 targets we have detected 4 systems
that have companions. Since our survey is flux limited we need to
correct for our bias towards detecting binaries that ``leak'' into our
sample from further distances. Our selection of $Ks<12$ leads to
incompleteness of single stars past $D \sim 22$ pc. Our detected
binaries have an incompleteness past 25 pc. Therefore, we are probing
1.46 times more volume with the brighter binaries compared to the
single (hence fainter) M8-M9 stars. Hence, the corrected binary
frequency is $4/20/1.46 = 14\%$.

Of course there are other selection effects due to the instrumental
PSF which prevents detection of very faint companions very close to the
primaries. We were only sensitive to companions of $\Delta K^\prime
\sim 1 mag$ at $0.13-0.17\arcsec$ separations. Much fainter companions
($\Delta K^\prime
\sim 5 mag$) could be detected at slightly wider ($\sim 0.25\arcsec$)
separations, and very low mass companions ($\Delta H \sim 10
mag$) could be detected at $\sim 1\arcsec$ separations. Therefore, we likely
are not detecting faint companions in the separation
range of $0.13-0.17\arcsec$. However, if we assume that the mass ratio
distribution ($q$) for M8-M9 stars is similar to that of M0-M6 binaries
(e.g. as least as many binaries with $\Delta K > 1.0 $ as $\Delta K < 1.0$ mag
\citep{fis92}), then based on our detection of 3 systems with $\Delta K
< 1 $ mag with separations of $0.13-0.17\arcsec$ we likely missed at
least $\sim 3$ other systems with $\Delta K > 1 $ mag with separations
in the range $0.13-0.17\arcsec$. Based on this assumption about the
mass ratio distribution there should be $\sim6$
binaries from $0.13-0.17\arcsec$ when correcting for our instrumental
insensitivity. Therefore, the total count should be 7 systems. Therefore,
the corrected M8-M9 binary frequency would be $7/20/1.46 = 24\%$ for
separations $>0.13\arcsec$. Hence we have a range of possible binary frequencies from $14\%$ (if there are no binaries with $\Delta K > 1 $ mag with separations of $0.13-0.17\arcsec$) up to $24\%$ if the $q$ distribution is similar to M0-M6 stars and we correct for insensitivity. {\it In any case,
we can state that for systems with separations greater than 
$sep >3$ AU (or $P >15$ yr) the M8-M9 binary frequency is likely within the range $14-24\%$}.

From \cite{fis92}, it appears that our M8-M9 binary fraction range of
$14-24\%$ is likely consistent with the $23\pm 5\%$ measured for more
massive M stars (M0-M6) over the same separation/period range ($P>15
yr$). However, the M8-M9 binaries are very different from the M stars
in the distribution of their semi-major axis. The M8-M9 binaries
appear to peak at $sep\sim4$ AU which is significantly tighter than
the $\sim$30 AU peak of both the G and M star binary
distributions. This cannot be a selection effect since we are highly
sensitive to all M8-M9 binaries with $sep>20-600$ AU (even those with
$\Delta K>10$ mag). {\it Therefore, we may conclude that M8-M9 stars
likely have similar binary fractions as G and M stars, but they have
significantly smaller semi-major axes on average}.

More observations of such systems will be required to see if these
trends hold over bigger samples. It is interesting to note that in
\cite{rei01a} a survey of 8 L binaries finds a similar binary frequency of 20\%
and a maximum separation of only 9 AU. Therefore it appears both M8-M9
and L binaries may have similar binary frequencies and smaller separations
than their more massive M and G counterparts.



\acknowledgements

The Hokupa'a AO observations were supported by
the University of Hawaii AO group. 
(D. Potter, O. Guyon, P. Badouz and
A. Stockton). Support for Hokupa'a comes from the National Science
Foundation. 
We thank the anonymous referee for comments that led to an improved analysis of the data and an all round better paper. 
LMC acknowledges support by the AFOSR under
F49620-00-1-0294. 
We would also like to send a big {\it
mahalo nui} to the Gemini operations staff (especially 
Simon Chan) for a flawless night.
These results were based on observations obtained at the Gemini
Observatory, which is operated by the Association of Universities for
Research in Astronomy, Inc., under a cooperative agreement with the
NSF on behalf of the Gemini partnership: the National Science
Foundation (United States), the Particle Physics and Astronomy
Research Council (United Kingdom), the National Research Council
(Canada), CONICYT (Chile), the Australian Research Council
(Australia), CNPq (Brazil) and CONICET (Argentina).





\clearpage
\begin{deluxetable}{lllllllll}
\tabletypesize{\scriptsize}
\tablecaption{The new binary systems observed Sept 22, 2001 \label{tbl-1}}
\tablewidth{0pt}
\tablehead{
\colhead{System} &
\colhead{$\Delta J$} &
\colhead{$\Delta H$} &
\colhead{$\Delta K^{\prime}$} &
\colhead{$\Delta K$} &
\colhead{Sep. ($\arcsec$)} &
\colhead{PA} &
\colhead{Age (Gyr)} &
\colhead{Est. D (pc)\tablenotemark{a}}
}
\startdata
2M1426\tablenotemark{b} & $0.78\pm 0.05$ & $0.70\pm 0.05$ &$0.65\pm 0.10$ &$0.57\pm 0.14$ & $0.152\pm0.006$ & $344.1\pm0.7^{\circ}$ & $0.8_{-0.2}^{+6.7}$\tablenotemark{b} & $23.6\pm 6.0$\\
2M2140 & $0.77\pm 0.05$ & $0.73\pm 0.04$ &$0.75\pm 0.04$ &$0.76\pm 0.13$ & $0.155\pm0.005$ & $134.3\pm0.5^{\circ}$ & $3.0_{-2.4}^{+4.5}$ & $23.9\pm 6.0$\\
2M2206 &$0.17\pm 0.04$ & $0.08\pm 0.04$ &$0.08\pm 0.03$ &$0.08\pm 0.14$ & $0.168\pm0.007$ & $68.2\pm0.5^{\circ}$ & $3.0_{-2.4}^{+4.5}$ & $24.68\pm 6.8$\\
2M2331 &$2.78\pm 0.04$ & $2.64\pm 0.05$ &$2.44\pm 0.03$ &$2.38\pm 0.16$ & $0.573\pm0.008$ & $302.6\pm0.4^{\circ}$ & $5.0_{-4.4}^{+2.5}$ & $25.2\pm 6.8$\\
\enddata
\tablenotetext{a}{photometric distances of the primaries calculated from M8-M9 relation of $M_K=7.593+2.25(J-Ks)$ from \cite{giz00})}
\tablenotetext{b}{2M1426 observations made on June 20, 2001 \cite{clo02}; the young age of 2M1426 is motivated in \cite{clo02}}
\end{deluxetable}

\clearpage
\begin{deluxetable}{llllllllll}
\tabletypesize{\scriptsize}
\tablecaption{Summary of the new binaries' A \& B components \label{tbl-2}}
\tablewidth{0pt}
\tablehead{
\colhead{Name} &
\colhead{$J$} &
\colhead{$H$} &
\colhead{$Ks$} &
\colhead{$K$} &
\colhead{$J-Ks$} &
\colhead{SpT\tablenotemark{a}}&
\colhead{Est. Mass\tablenotemark{b}} &
\colhead{Sep. (AU)} &
\colhead{P (yr)\tablenotemark{c}}
}
\startdata
2M1426A &$13.36\pm 0.06$ &$12.63\pm 0.05$ &$12.20\pm 0.07$ &$12.07\pm 0.08$ & $1.16\pm0.12$& M8.5 & $0.083_{-0.014}^{+0.010}$& $3.6\pm 0.9$& $17_{-7}^{+10}$\\
2M1426B &$14.13\pm 0.06$ &$13.34\pm 0.10$ &$12.80\pm 0.14$ &$12.64\pm 0.14$ & $1.33\pm0.12$& L1 & $0.075_{-0.020}^{+0.009}$ & & \\

2M2140A &$13.37\pm 0.06$ &$12.72\pm 0.04$ &$12.22\pm 0.04$ &$12.07\pm 0.09$ & $1.15\pm0.12$& M8.5 & $0.087_{-0.017}^{+0.008}$& $3.7\pm 0.9$& $18_{-7}^{+10}$\\
2M2140B & $14.15\pm 0.06$ & $13.44\pm 0.04$ &$12.97\pm 0.04$ &$12.83\pm 0.09$ & $1.19 \pm 0.14$& L0 & $0.075_{-0.018}^{+0.007}$ & &  \\

2M2206A &$13.12\pm 0.06$ &$12.46\pm 0.06$ &$12.06\pm 0.05$ &$11.94\pm 0.08$ & $1.06\pm0.12$& M8.0 & $0.090_{-0.014}^{+0.008}$& $4.1\pm 1.1$& $20_{-7}^{+9}$\\
2M2206B & $13.27\pm 0.06$ & $12.54\pm 0.06$ &$12.14\pm 0.05$ &$12.02\pm 0.08$ & $1.13 \pm 0.14$& M8.5 & $0.088_{-0.014}^{+0.008}$ & &  \\

2M2331A &$13.08\pm 0.04$ &$12.38\pm 0.04$ &$12.04\pm 0.04$ &$11.94\pm 0.07$ & $1.04\pm0.12$& M8.0 & $0.091_{-0.013}^{+0.008}$& $14.4\pm 3.9$& $139_{-57}^{+86}$\\
2M2331B & $15.86\pm 0.06$ & $15.03\pm 0.06$ &$14.48\pm 0.06$ &$14.32\pm 0.10$ & $1.38 \pm 0.14$& L3 & $0.062_{-0.020}^{+0.010}$ & &  \\

\enddata
\tablenotetext{a}{Spectral types estimated from J-Ks colors and calibrations of \cite{rei01b} with $\pm1.5$ spectral subclasses of error in these estimates.}
\tablenotetext{b}{Masses (in solar units) from the models of \cite{cha01} --see Figure \ref{fig2}}
\tablenotetext{c}{Periods estimated assuming face-on circular orbits} 
\end{deluxetable}

\clearpage

\clearpage

\begin{thebibliography}{}

\bibitem[Burrows et al.(2000)]{bur00}Burrows, A. et al. 
2000, Protostars and Planets IV 
p. 1339
\bibitem[Caloi et al.(1999)]{cal99}Caloi, V., et al. 
1999, A\&A, 351, 925.
\bibitem[Chabrier et al.(2000)]{cha00}Chabrier, G. et al. 
2000, \apj, 542, 464 
\bibitem[Chabrier et al.(2001)]{cha01}Chabrier, G. et al. 
2001, \apj, submitted 
\bibitem[Close et al.(2002)]{clo02}Close, L.M. et. al. 2002, \apj, in press (astro-ph/0110669).
\bibitem[Close et al.(1998)]{clo98}Close, L.M.
1998, Proc. SPIE Vol. 3353, p. 406-416.
\bibitem[Close(2000)]{clo00}Close, L. M. 2000, Proc. SPIE Vol. 4007, p758-772.
\bibitem[Duquennoy \& Mayor(1991)]{duq91}Duquennoy, A., Mayor, M. 1991, \aap 248, 485
\bibitem[Fischer \& Marcy(1992)]{fis92}Fischer, D. A., Marcy, G. W. 1992, \apj, 396, 178
\bibitem[Graves et al.(1998)]{gra98}Graves, J.E., Northcott, M.J., Roddier, F.J., Roddier, C.A., Close, L.M. 1988, Proc. SPIE Vol. 3353, p. 34-43.
\bibitem[Gizis et al.(2000)]{giz00}Gizis, J.E. et al. 
2000, \aj, 120, 1085
\bibitem[Hodapp et al.(1996)]{hod96}Hodapp, K.-W. et al. 
1996, New Astronomy, 1, 177
\bibitem[Kenworthy et al.(2001)]{ken01}Kenworthy, M., et al. 
2001, \apj, 554, L67 
\bibitem[Lane et al.(2001)]{lan01}Lane, B.F. et al. 
2001, \apj, in press
\bibitem[Mart\'\i n, Brandner, \& Basri(1999)]{mar99}Mart\'\i n, E. L., Brandner, W., Basri, G. 1999, Science, 283, 5408, 1718
\bibitem[Nakajima et al.(1995)]{nak95}Nakajima, T., et al.
1995, \nat, 378, 463
\bibitem[Reid et al.(2001b)]{rei01b}Reid, I. N. et al. 
2001b, \aj, 121, 1710
\bibitem[Reid et al.(2001a)]{rei01a}Reid, I. N. et al. 
2001a, \aj, 121, 489
\bibitem[Simon, Close \& Beck (1999)]{sim99} Simon, M., Close, L.M., \& Beck, T. 1999, \aj, 117, 1375 
\bibitem[Stetson(1987)]{ste87}Stetson, P. B. 1987, \pasp, 99, 191
\bibitem[Wainscoat \& Cowie (1992)]{wai92} Wainscoat R. J., \& Cowie, L.L. 1992, \aj 103, 332.

\end{thebibliography}
\end{document}